**Tuning the piezoelectric and mechanical properties of the AlN system via alloying with YN and BN**

Sukriti Manna[1], Geoff L. Brennecka[2], Vladan Stevanović[2] and Cristian V. Ciobanu[1*]

[1]*Dept. of Mechanical Engineering, Colorado School of Mines, Golden, Colorado 80401*

[2]*Dept. of Metallurgical and Materials Engineering, Colorado School of Mines, Golden, Colorado 80401*

**Abstract:** Recent advances in microelectromechanical systems often require multifunctional materials, which are designed so as to optimize more than one property. Using density functional theory calculations for alloyed nitride systems, we illustrate how co-alloying a piezoelectric material (AlN) with different nitrides helps tune both its piezoelectric and mechanical properties simultaneously. Wurtzite AlN-YN alloys display increased piezoelectric response with YN concentration, accompanied by mechanical softening along the crystallographic *c* direction. Both effects increase the electromechanical coupling coefficients relevant for transducers and actuators. Resonator applications, however, require superior stiffness, thus leading to the need to decouple the increased piezoelectric response from a softened lattice. We show that co-alloying of AlN with YN and BN results in improved elastic properties while retaining some of the piezoelectric enhancements from YN alloying. This finding may lead to new avenues for tuning the design properties of piezoelectrics through composition-property maps.

*Keywords: piezoelectricity, electromechanical coupling, density functional theory, co-alloying*

___________________________________

\* To whom correspondence may be addressed, electronic mail: cciobanu@mines.edu

**INTRODUCTION**

Recent technological advances in microelectromechanical systems require materials to be more efficiently designed, often leading to optimization of more than one property.[1, 2] An electromechanical (piezoelectric) material requires the best piezoelectric coefficient, while materials for damping high mechanical loads require the high stiffness moduli.[3, 4] The high stiffness value is also associated with thermal stability, enabling the use of these materials over a large temperature range.[5] Pseudo-binary alloys, i.e. solid solutions between two compounds, have been extensively investigated for use in electromechanical applications.[6-10] Most of these alloys have large miscibility gaps, which makes them less appealing for applications where mechanical stability is an important figure of merit. The presence of miscibility gaps is not particularly restrictive, since current non-equilibrium processing techniques allow for fabricating pseudo-binary alloys in which one can seek a given functionality by, e.g., changing composition (often through the miscibility gaps) so as to optimize piezoelectric coefficients, stiffness, electromechanical coupling constants, or electronic properties. However, achieving optimal or desired values for more than one such property is challenging since optimizing one property often changes another towards undesirable values. In such cases, more degrees of freedom may help create a framework in which advanced materials can be optimized for multiple functionalities. In this article, we focus on a piezoelectric system to show that alloying with more than one compound enables us to tune two properties, specifically, the piezoelectric coefficient and the mechanical stiffness.

With the discovery of the anomalously large increase in the piezoelectric moduli of wurtzite AlN when alloyed with ScN[11] or YN,[9, 12] AlN-based alloys have received a lot of attention as materials for improved telecommunications, sensors, and surface acoustic wave devices.[13, 14] The origin of the enhanced piezoelectricity was revealed to be an intrinsic effect of alloying, as opposed to textural or microstructural effects,[6] with the obvious implication that controlling the piezoelectric enhancement can be done mainly by increasing composition of YN or ScN. However, such alloying softens the material significantly,[6] which actually makes the material *less* attractive for resonant applications for which $k^2Q$ is a common figure of merit[15] wherein $k$ is an electromechanical coupling coefficient and $Q$ is a quality factor that is inversely proportional to mechanical dissipation.[15] It is therefore desirable to control the mechanical properties as well, which could in principle be achieved by alloying with other species.

Herein, we focus on AlN-YN alloys and advance the premise that further alloying this system with BN can increase the elastic constants of the material, thus leading to new avenues for piezoelectric design based on precise control YN and BN compositions. The choice for YN is motivated by the scarcity of literature reports on AlN-YN alloys,[12, 16] and by the fact that Y may be a less expensive alternative for Sc. While the elastic stiffening of AlN-YN upon alloying with BN would be enabled by the presence of shorter and stiffer B−N bonds, the effect of alloying with BN on the piezoelectric properties is presently unknown. Using density functional theory (DFT), we have found that the presence of BN, by itself, indeed leads to increases in the relevant elastic constant $C_{33}$, but does not change the piezoelectric coefficient $e_{33}$ significantly. Furthermore, we have uncovered the combined effect of alloying with YN and BN on the properties of the AlN system for total alloying concentrations up to 50 at. %. The addition of BN to AlN-YN alloys counters to some extent the piezoelectric enhancement obtained via YN alloying, but increases the elastic modulus. Overall, the alloying of AlN with BN and YN leads to a high degree of control over the piezoelectric and elastic properties of the resulting alloy, and therefore over the coupling coefficients and various figures of merit commonly used in piezoelectric device design.

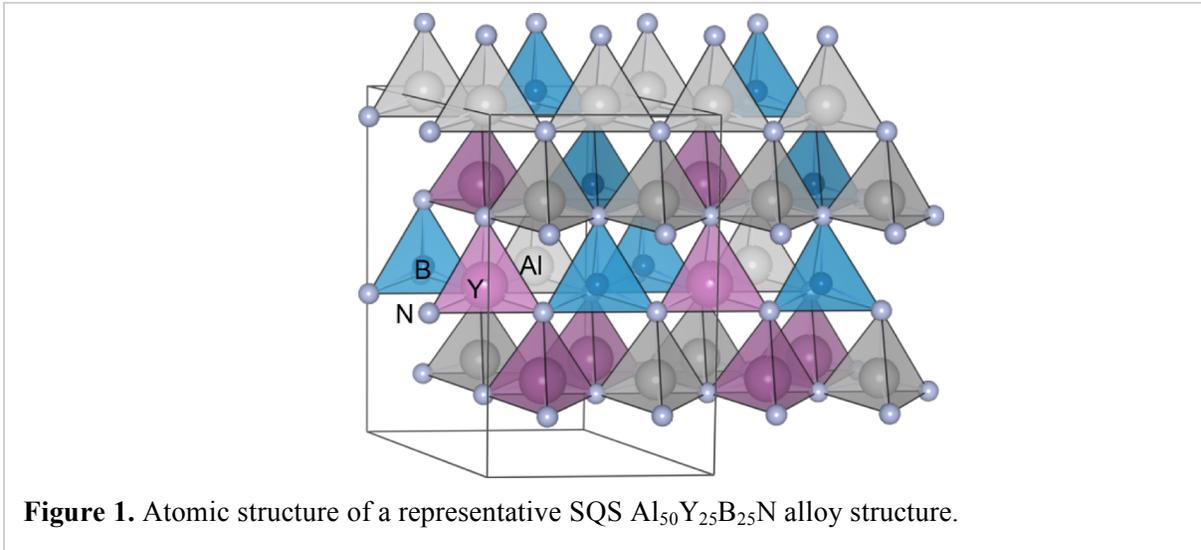

**Figure 1.** Atomic structure of a representative SQS $Al_{50}Y_{25}B_{25}N$ alloy structure.

**METHODS**

To assess the changes in elastic and piezoelectric properties driven by different levels of YN and BN in $Al_{1-x-y}Y_xB_yN$ alloys, we first performed DFT relaxations of atomic positions and lattice constants using the VASP package,[17] with projector augmented waves[18] in the generalized gradient approximation using the Perdew-Burke-Ernzerhof (PBE) exchange-correlation functional.[19] Out of caution more than necessity, we introduced an on-site Coulomb interaction (parameter $U = 3$ eV) for the yttrium $d$ state since this state is explicit in the pseudopotential. When compared with calculations done with plain PBE ($U = 0$ eV), the results for lattice constants are nearly identical, while the piezo and stiffness tensor component vary by less than 2%. The parent structure is wurtzite AlN, in which Al is substituted with Y and B in desired amounts. Supercells with 32 atoms are simulated using special quasirandom structures (SQS)[20] (Fig. 1), whose pair correlation functions are the same as those of random alloys up to second-nearest neighbors. A 5×5×3 Monkhorst–Pack[21, 22] $k$-point mesh was used to sample the Brillouin zone, resulting in 38 irreducible $k$-points. A plane wave cutoff energy of 500 eV was used; the energy tolerance stopping criteria were $10^{-8}$ and $10^{-7}$ for the electronic and ionic convergence, respectively. These criteria were obtained by progressive tightening of the convergence until the relaxed SQS supercells stopped changing. This resulted in smoother curves for lattice parameters and properties, in which the only source of noise remained the use of small number of atoms coupled with a limited number of individual SQS supercells at each concentration (refer to Fig. SM-1 in Supplementary Material for an assessment of the dispersion of individual SQS values). Due to the variability of atomic-scale environments in the SQS supercells,[8] we performed five calculations at each composition and used them to average the properties, including the lattice constants. For the separate alloying only with YN or with BN, the lattice constants $a$ and $c$ are plotted in Fig. 2(a,b) as functions of the cation concentration. Alloying with YN increases the lattice constants, while alloying with BN alone decreases them: these trends are simply due to the larger (smaller) cation radius of Y (B) compared to that of Al in AlN. The $c/a$ ratio is more sensitive to doping for the case of Y compared to BN [Fig. 2(c)].

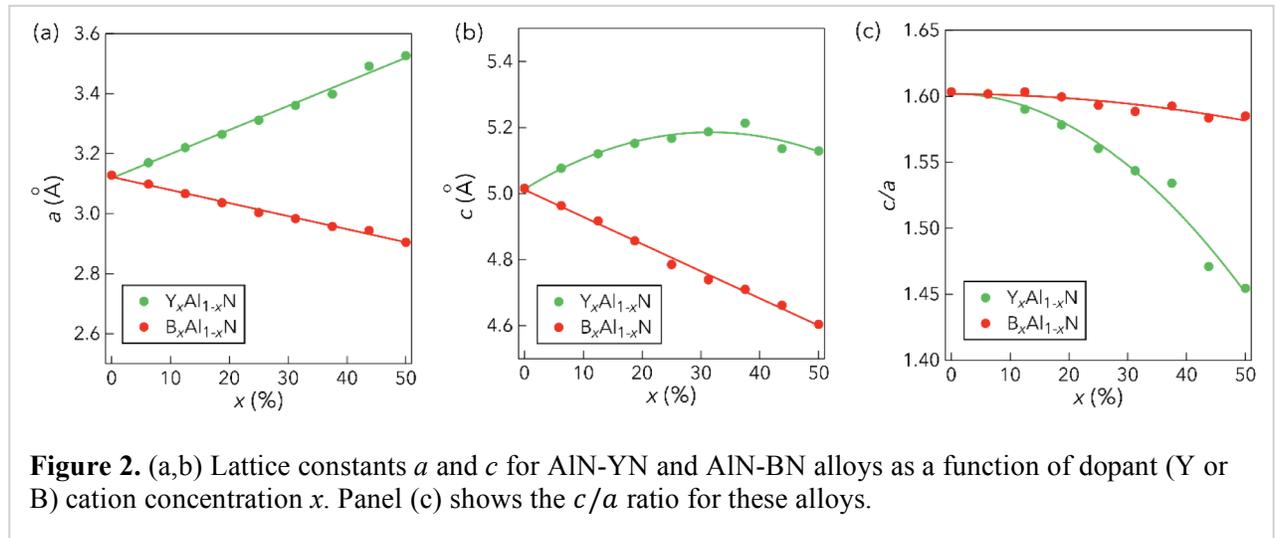

**Figure 2.** (a,b) Lattice constants $a$ and $c$ for AlN-YN and AlN-BN alloys as a function of dopant (Y or B) cation concentration $x$. Panel (c) shows the $c/a$ ratio for these alloys.

Our DFT calculations of mixing enthalpies for alloy wurtzite and rocksalt phases of AlN-YN revealed that wurtzite is more stable than rocksalt for up to ~72% Y content, above which rocksalt becomes more stable [Fig. 3(a)]. The mixing enthalpies in Fig. 3(a) are positive, meaning that mixing is not spontaneous. From the mixing enthalpies and combinatorial expression for mixing enthalpy in this pseudo-binary alloy, we have obtained the temperature-composition phase diagram in a manner similar to that described by, e.g., Burton et al. for other wurtzite systems.[23] This phase diagram is shown in Fig. 3b, in which the temperature axis is rather qualitative since we have not considered lattice vibrations in deriving it; it is similar to the one that would be obtained in the presence of vibrations.[23] As would be expected, the diagram shows a miscibility gap – the white space between the metastable regions in Fig. 3b. This means that compositions in that gap cannot be achieved under thermodynamic equilibrium. However, the current growth techniques (e.g. reactive dc magnetron co-sputtering)[24-26] rely on non-equilibrium processes to overcome the solubility limits[27-29] and create alloys with better control over properties through control of composition.

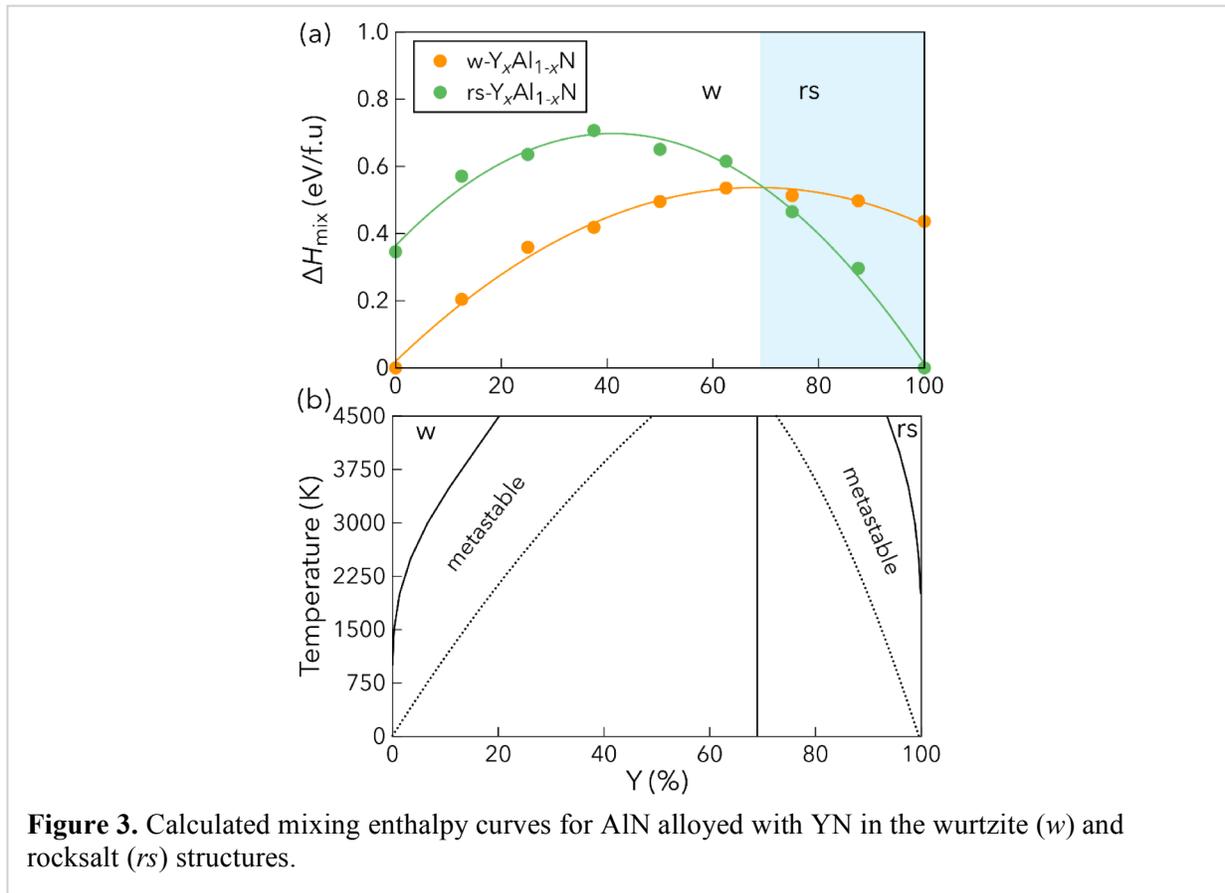

**Figure 3.** Calculated mixing enthalpy curves for AlN alloyed with YN in the wurtzite (*w*) and rocksalt (*rs*) structures.

## RESULTS AND DISCUSSION

Focusing now on describing the piezoelectric properties of the alloys, we note that the piezoelectric coefficient $e_{33}$ is evaluated at each Y ($x$) and B ($y$) composition via

$$e_{33} = e_{33}^{clamped-ion} + \frac{4eZ^*}{\sqrt{3}\,a_0^2}\frac{du}{d\varepsilon} \qquad (1)$$

where $e_{33}^{clamped-ion}$ is the clamped ion contribution to $e_{33}$, $e$ is the electron charge, $Z^*$ is the axial component of the dynamical Born effective charge tensor, $a_0$ is the equilibrium lattice constant, and $du/d\varepsilon$ is the sensitivity of the internal parameter $u$ to strains along the crystallographic *c*-axis.[23, 24] The elastic constants were computed by finite differences between the energies of the lattices subjected to small applied strains.[30] Figure 4 shows the effects of separate BN and YN alloying on the properties of AlN. As seen in Fig. 4(a) the piezoelectric coefficient $e_{33}$ increases with Y concentration up to 50%, while the elastic constant $C_{33}$ decreases; this is consistent with previous works on the similar AlN-ScN system.[6, 8] The physical origin of this effect arises from the larger radius of the Y ion, which changes the internal parameter $u$ of the alloyed wurtzite structure and its sensitivity to strain, $du/d\varepsilon$ [refer to Eq. (1)]. The value of $e_{33}$ that can be achieved by alloying with 50% Y is doubled compared to that of pure AlN (Fig. 4). Figure 4(b) shows that alloying with BN alone increases the elastic modulus, which is consistent to another report.[31] Interestingly, we also note that substituting B for Al does not alter the piezoelectric coefficient significantly up to concentrations of 50% B.

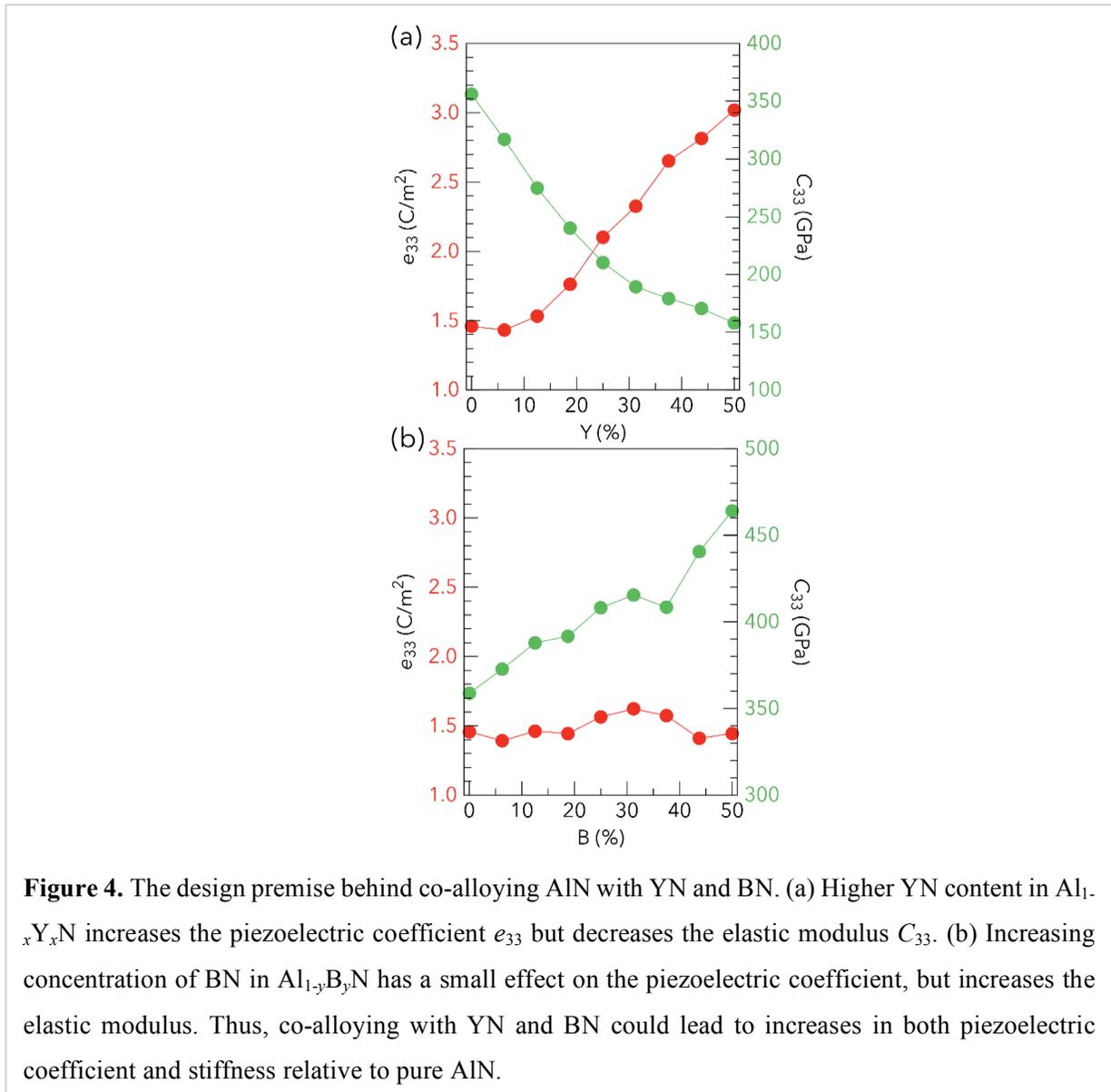

**Figure 4.** The design premise behind co-alloying AlN with YN and BN. (a) Higher YN content in Al$_{1-x}$Y$_x$N increases the piezoelectric coefficient $e_{33}$ but decreases the elastic modulus $C_{33}$. (b) Increasing concentration of BN in Al$_{1-y}$B$_y$N has a small effect on the piezoelectric coefficient, but increases the elastic modulus. Thus, co-alloying with YN and BN could lead to increases in both piezoelectric coefficient and stiffness relative to pure AlN.

When AlN-YN films are deposited, there is usually strain in the lattice stemming from the mismatch with the substrate, changing film texture with alloy concentration, or from insufficient annealing. It is therefore relevant to assess what the effect of this strain is on the piezoelectric coefficient, so we have re-done the DFT calculations for the situations were strain is applied in plane (along the *a* and *b* directions) or out of plane (along the *c* direction). The results of these calculations are displayed in Fig. 5: $e_{33}$ can increase up to about 3.75 C/m² (i.e., 2.5 times that of pure AlN) either by alloying with 50% Y and applying 2% tension along the *a* axis [Fig. 5(a)], or by alloying with 50% Y and applying 2% compression along the *c* axis [Fig. 5(b)]. The opposite trends of $e_{33}$ with strain along different directions (Fig. 5) suggest that, for homogeneous and isotropic strain in the AlN-YN lattice, the effects of strain along *c* and *a* directions may cancel each other at least partially and thus not affect the piezoelectric coefficient significantly. This is important when alloying with a secondary species, such as BN, which would amount to roughly isotropic strain when B is (nearly) homogenously doped throughout the AlN-YN lattice.

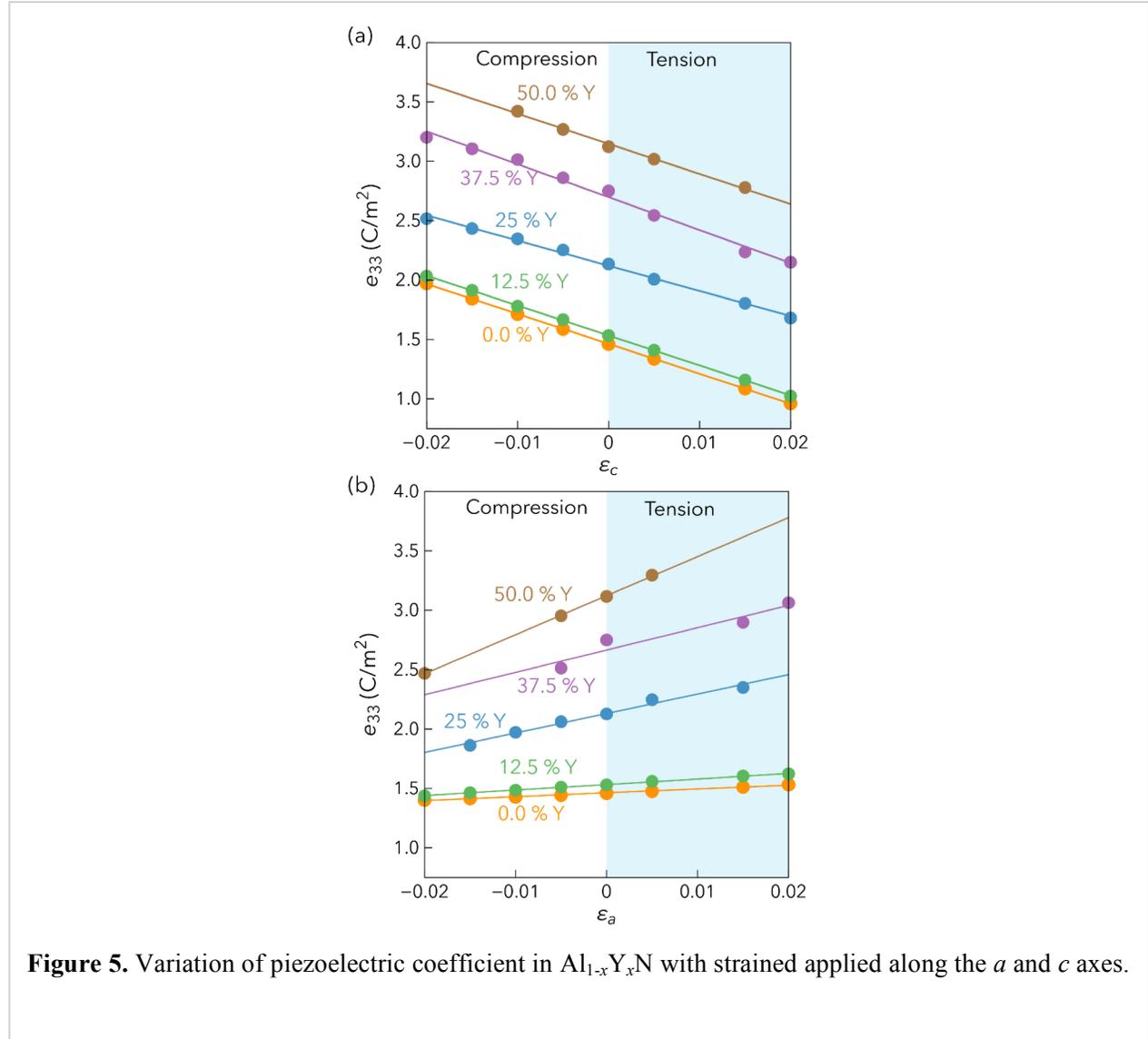

**Figure 5.** Variation of piezoelectric coefficient in $Al_{1-x}Y_xN$ with strained applied along the *a* and *c* axes.

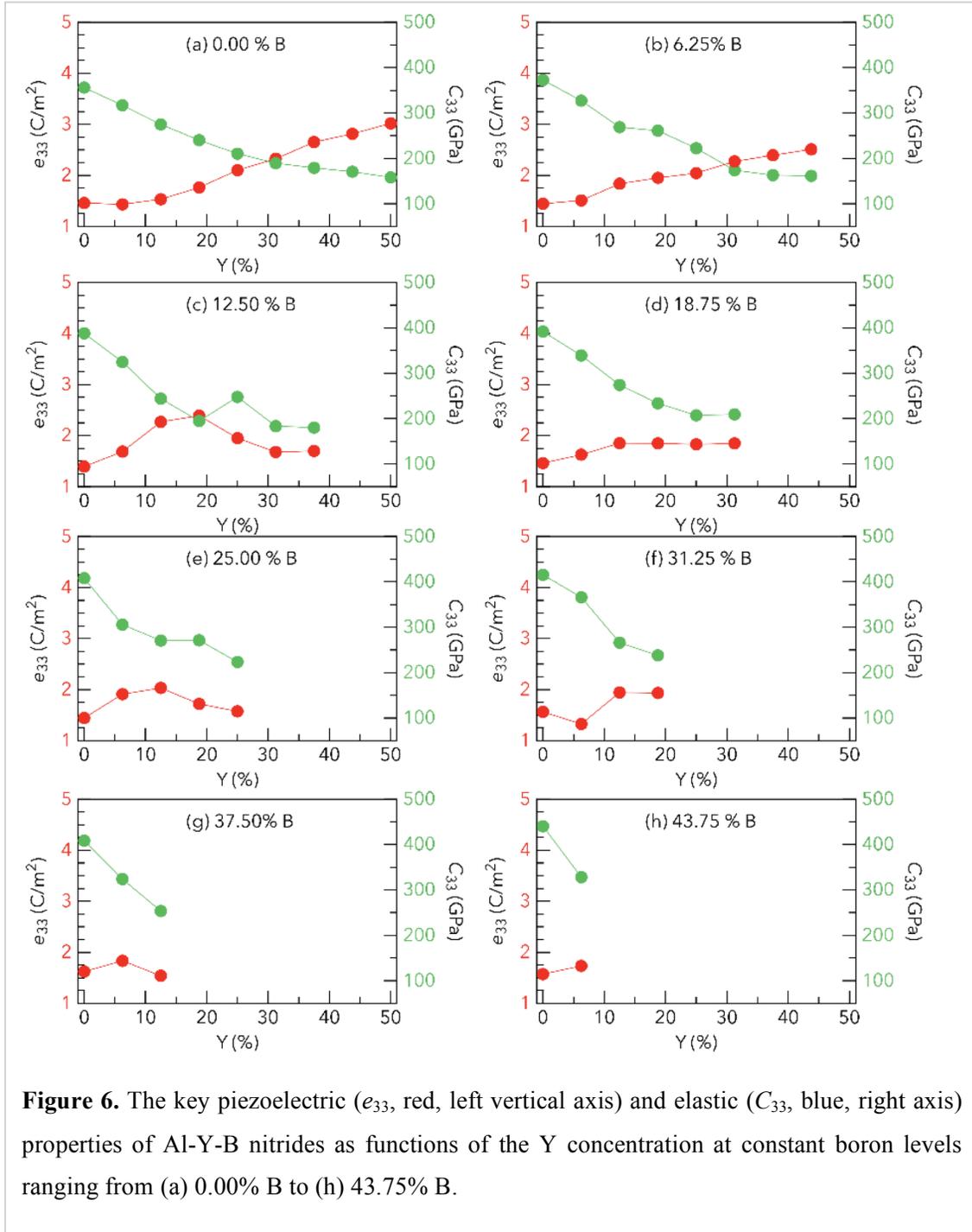

**Figure 6.** The key piezoelectric ($e_{33}$, red, left vertical axis) and elastic ($C_{33}$, blue, right axis) properties of Al-Y-B nitrides as functions of the Y concentration at constant boron levels ranging from (a) 0.00% B to (h) 43.75% B.

Based on these results, we have set out to analyze how co-alloying with both YN and BN changes properties, aiming to find composition regimes where the piezoelectric coefficient increases while the elastic modulus does not decrease to the same extent as in the case of alloying with Y alone. Figure 6 shows the variation of $e_{33}$ and $C_{33}$ with the Y concentration $x$, at a fixed level of B doping, $y$ (such that $x+y \leq 50\%$). The data points in Fig. 6 represent averages over SQS configurations. Other properties (i.e., $e_{31}$ and $C_{11}$) are shown as functions of $x$ and $y$ in Fig. SM-2.

Our starting premise that the presence of boron can mitigate or control the decrease in elastic modulus $C_{33}$ brought about by Y alloying – while still retaining some increase of the piezoelectric coefficient – is borne out by the results in Fig. 6. For example, going from 0% B in $Al_{50}Y_{50}N$ to 18.75% B concentration in $Al_{50}Y_{31.25}B_{18.75}N$ results in a ~50 MPa increase of the elastic constant $C_{33}$ with ~33% decrease in the piezoelectric coefficient. For ease of reference, some of the results shown in Fig. 6 are also tabulated in Table I. BN alloying has a strong effect on mechanical properties (Table I): increasing BN content from 31.25% to 50% at the expense of YN concentration nearly doubles the elastic modulus (from 238 GPa to 470 GPa).

**Table I.** Comparison between the properties of some of the Al-Y-B nitrides computed in this work and those of three commercial piezoelectrics.

| Material | $e_{33}$ (C/m$^2$) | $C_{33}$ (GPa) | $d_{33}$ (pC/N) | $k_{33}^2$ |
|---|---|---|---|---|
| PZT-7A, Ref.[32] | 9.50 | 131 | 153 | 0.67 |
| PZT-4, Ref.[32] | 13.84 | 113 | 225 | 0.35 |
| $K_{0.5}Na_{0.5}NbO_3$, Ref.[33, 34] | 4.14 | 104 | 80, 130 | 0.51 |
| *this work:* | | | | |
| AlN | 1.46 | 353 | 5 | 0.06 |
| $Al_{50}Y_{50}N$ | 3.02 | 158 | 38 | 0.37 |
| $Al_{50}Y_{43.75}B_{6.25}N$ | 2.51 | 161 | 22 | 0.29 |
| $Al_{50}Y_{37.50}B_{12.50}N$ | 1.70 | 180 | 13 | 0.14 |
| $Al_{50}Y_{31.25}B_{18.75}N$ | 1.85 | 209 | 13 | 0.14 |
| $Al_{50}Y_{25}B_{25}N$ | 1.58 | 223 | 13 | 0.11 |
| $Al_{50}Y_{18.75}B_{31.25}N$ | 1.93 | 238 | 13 | 0.14 |
| $Al_{50}Y_{12.5}B_{37.50}N$ | 1.54 | 253 | 9 | 0.12 |
| $Al_{50}Y_{6.25}B_{43.75}N$ | 1.73 | 328 | 8 | 0.08 |
| $Al_{50}B_{50}N$ | 1.34 | 470 | 3 | 0.04 |

Next, we discuss the electromechanical coupling coefficients resulting from co-alloying of AlN with YN and BN. These coefficients are key parameters in designing transducers for energy harvesting and sensing applications.[5, 35] Applications such as pressure sensors, accelerometers, and gyroscopes commonly require the piezoelectric medium to operate in the longitudinal length mode,[36] for which the relevant coupling coefficient is

$$k_{33}^2 = \frac{e_{33}^2}{\varepsilon_{33} C_{33} + e_{33}^2}$$

where $\varepsilon_{33}$ is the 33 component of the dielectric tensor. The static dielectric constants vary by about 10-15 % from their value at $x$=12.5% (separate calculation) for the entire range of equimolar Y-B doping (i.e., $x \leq 25\%$ in $Al_{1-2x}Y_xB_xN$), so a constant value of $\varepsilon_{33} = 11.0$ was assumed in computing the coupling coefficients. High $k_{33}^2$ coupling factor corresponds to transducers with better axial resolution, broader bandwidth, and higher sensitivity.[37] Actuators based on cantilevers operate in the transverse length mode

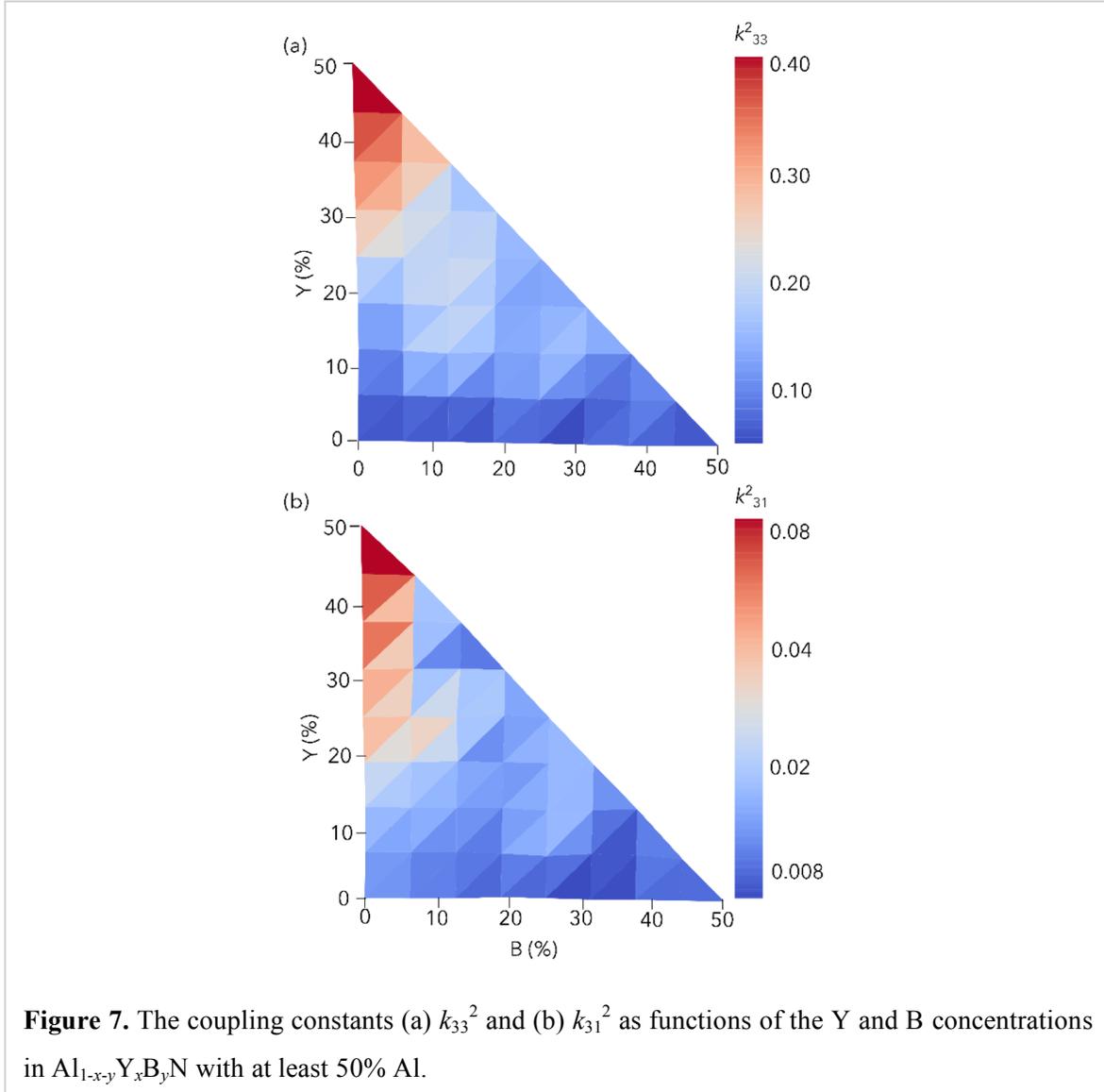

**Figure 7.** The coupling constants (a) $k_{33}^2$ and (b) $k_{31}^2$ as functions of the Y and B concentrations in $Al_{1-x-y}Y_xB_yN$ with at least 50% Al.

(bending),[35, 38] for which the relevant coupling coefficient is

$$k_{31}^2 = \frac{e_{31}^2}{\varepsilon_{33} C_{11} + e_{31}^2},$$

where $e_{31}$ and $C_{11}$ are components of the piezoelectric and elastic tensor, respectively. The two coupling coefficients are plotted in Fig. 7 (a,b), for the range of Y and B concentrations considered. High coupling constants require large piezoelectric coefficients and low elastic moduli. Increasing the amount of Y in the system increases the piezoelectric and coupling coefficients but softens the system elastically,[6] which is true for both coupling modes addressed here.

The two coupling coefficients exhibit similar trends (Fig. 7), with larger values achieved for higher Y concentrations. However, variations of $k_{31}^2$ are significantly smaller, and larger values can also be achieved using equiatomic Y:B alloys. With the piezoelectric and elastic tensor components obtained from DFT calculations, it is useful to compare some of the properties of the co-alloyed AlN system with those of known piezoelectrics. Table I, which also includes the longitudinal piezoelectric modulus $d_{33}$, shows the properties of three well-known piezoelectrics in comparison with a few selected alloy compositions computed here (for completeness, piezoelectric moduli are also plotted in Fig. SM-3 for the entire range of compositions used). Direct inspection shows that AlN co-alloyed with Y and B can achieve parameters comparable to those of commercial piezoelectrics. More importantly, the simultaneous enhancement of $e_{33}$ and $C_{33}$ in comparison to the pure AlN values can be seen for boron levels of up 25%, illustrating that co-alloying with different species represents a valuable route for simultaneously engineering the piezoelectric and the elastic properties of a parent wurtzite structure such as that of AlN.

**CONCLUDING REMARKS**

In conclusion, we have shown that co-alloying of AlN with YN and BN leads to superior control of both piezoelectric and mechanical properties. The idea of introducing BN into the systems arose as a way to stiffen the material, which otherwise would have becomes increasingly softer with increased YN content. In the process of assessing the influence of BN and YN composition on properties, we have separately calculated the effect of straining the lattice in different directions, and found that enhancements of the piezoelectric coefficient can be made by applied strain as well: this strain effect could become important for certain substrates, or for electromechanical applications where strain develops in the structure during operation.

The results presented here could carry over to other parent phases and alloying phases as well, provided that each alloying agent offers a clear advantage in terms of engineering one specific property. In our case, YN leads to increasing the piezoelectric coefficient and BN leads to increasing the stiffness. The use of both YN and BN offers intrinsic and simultaneous control over the mechanical and piezoelectric properties of the Al-Y-B-N system, even though the maximal value for each individual parameter can be achieved in the presence of only one dopant.

*Acknowledgment.* The authors gratefully acknowledge the support of the National Science Foundation through Grant No. DMREF-1534503. The calculations were performed using the high-performance computing facilities of the Golden Energy Computing Organization and the National Renewable Energy Laboratory.